\documentclass{article}
\usepackage{url}
\usepackage[a4paper, total={15cm, 24cm}]{geometry}
\usepackage{newtxtext,newtxmath}
\usepackage{titling}
\usepackage{graphicx}
\usepackage{subcaption}
\usepackage{amsmath}
\usepackage{microtype}
\title{Integrated RGB Beam Combiner in Al$_2$O$_3$ Visible Photonic Circuits with On-Chip Modulation for AR/VR Displays}
\author{Vahram Voskerchyan\textsuperscript{1} \\[1ex]
\text{v.voskerchyan@utwente.nl}}
\date{}

\begin{document}

\maketitle

\begin{center}
\end{center}

\vspace{7.5ex}

\textit{Aluminum oxide (Al$_2$O$_3$) is a promising material for visible-light integrated photonics due to its low optical loss and wide transparency window across the RGB spectrum. This work presents the design and experimental demonstration of an integrated RGB beam combiner for applications in AR/VR, holography, 3D displays, and autostereoscopic display systems. The device employs Mach--Zehnder modulators for individual color modulation and gratings for out-of-plane emission. The experimental results demonstrate independent RGB routing, far field beam combining, and thermo-optic modulation with an extinction ratio of up to 6.3~dB, highlighting the potential of Al$_2$O$_3$ photonic integrated circuits for compact dynamic color control at the pixel level.}

\section*{Introduction}

Integrated photonics has enabled compact, scalable optical systems for various applications. One area that benefits greatly from these advances is 3D display systems, particularly autostereoscopic displays. These displays generate 3D images without the need for special glasses, enabling glasses-free three-dimensional perception. Achieving such effects requires precise control of light, specifically, the ability to manipulate the intensity and color of light at the pixel level. Integrated photonics enables this control in a compact and scalable format, which is crucial for high-resolution and energy-efficient 3D displays \cite{notaros2024}.

Among the various materials used in integrated photonics, aluminum oxide stands out as particularly suitable for visible-light applications. It offers low optical loss and excellent transparency across the visible spectrum, making it ideal for devices that require the manipulation of red, green, and blue (RGB) light \cite{hendriks2021}. These three colors are the foundation of full-color displays, which are used in systems like augmented and virtual reality (AR/VR) and glassless 3D displays \cite{notaros2024}. By integrating the control of RGB light into a single platform, more efficient and compact display systems can be created, driving the next generation of visual technologies.

Recent developments have demonstrated various approaches for integrating RGB light on a chip \cite{nakao2020,SAKAMOTO,shi2025}. This paper explores a novel design for an RGB beam combiner on an aluminum oxide platform, which combines red (632 nm), green (520 nm), and blue (452 nm) optical channels. The design utilizes Mach--Zehnder modulators (MZMs) to control the intensity of each color channel. The modulated light is then routed to the corresponding RGB gratings that diffract the beams at a common emission angle, enabling overlap in the far field. The MZMs allow for dynamic tuning of each color, enabling fine control of the RGB signal serving as a building block for a pixel.

By leveraging aluminum oxide’s favorable optical properties, this RGB combiner offers a scalable and compact solution for 3D display technologies and AR. Its design is flexible, allowing for future enhancements such as the incorporation of an array of slanted gratings to achieve autostereoscopic effect \cite{raval2020} by directing the light at different angles without the need for phase tuning of each individual grating. The development of this RGB beam combiner could lead to smaller, more energy-efficient, and higher-performing displays for a wide range of applications, from entertainment to medical imaging, holography and beyond.

\section*{Concept}

\begin{figure}[htbp]
    \centering
    \begin{minipage}[b]{0.35\textwidth}
        \centering
        \includegraphics[width=\textwidth]{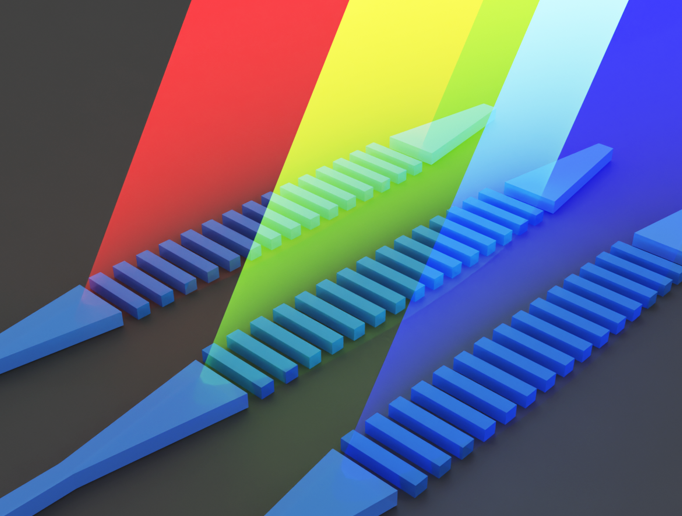}
        \text{(a)}
    \end{minipage}
    \hfill
    \begin{minipage}[b]{0.6\textwidth}
        \centering
        \includegraphics[width=\textwidth]{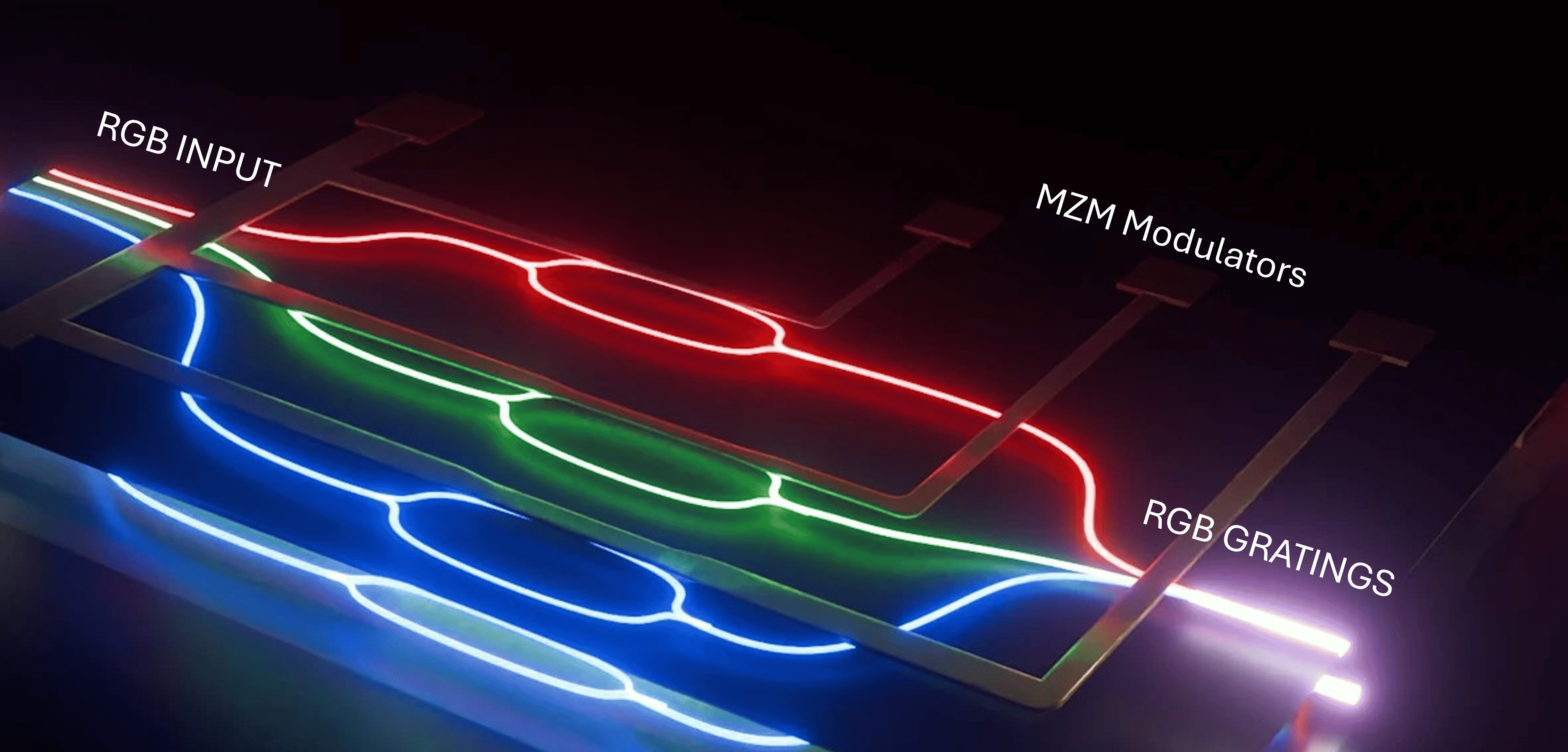}
        \text{(b)}
    \end{minipage}
    \caption{(a) Concept for a single RGB pixel in the proposed beam combiner design. (b) 3D visualization and schematic layout.}
    \label{fig:rgb_pixel_concept}
\end{figure}

Figure \ref{fig:rgb_pixel_concept}(a) illustrates the concept of a single RGB pixel in the proposed beam combiner design. Red (632 nm), green (520 nm), and blue (452 nm) light are coupled into individual Mach--Zehnder modulators (MZMs), which independently control the intensity of each color via phase modulation. This architecture enables precise intensity distribution to specific RGB gratings, which diffract each wavelength into the far field at a fixed, overlapping angle. By dynamically adjusting the color balance at the pixel level, the system enables the high-fidelity color reproduction required for next-generation AR/VR displays. This architecture was specifically chosen to navigate current foundry limitations by utilizing standard building blocks to realize a robust, simple-to-fabricate proof-of-concept for an integrated laser pixel.

\section*{Design and Simulations}

\begin{figure}[htbp]
    \centering
    \begin{subfigure}[b]{0.48\textwidth}
        \centering
        \includegraphics[width=\textwidth]{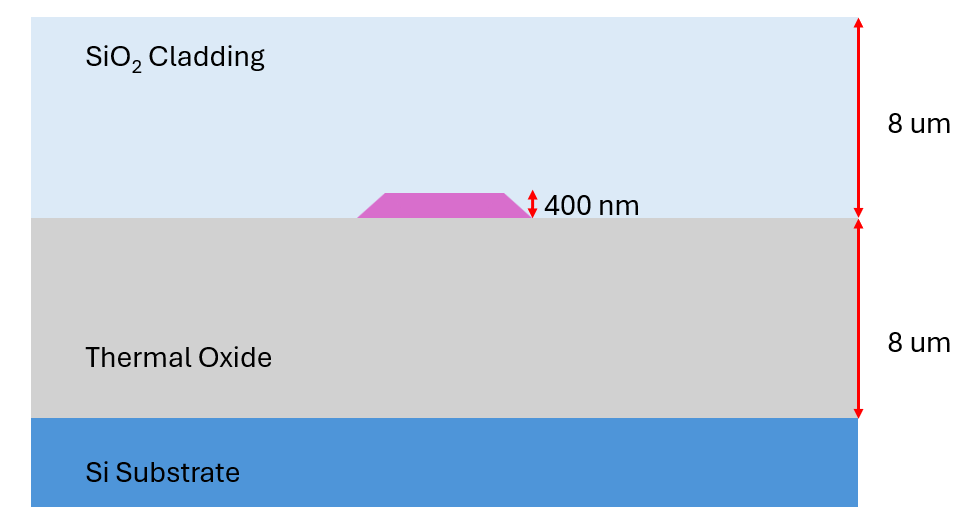}
        \caption{Layer stack of the fabricated device.}
        \label{fig:layerstack}
    \end{subfigure}
    \hfill
    \begin{subfigure}[b]{0.48\textwidth}
        \centering
        \includegraphics[width=\textwidth]{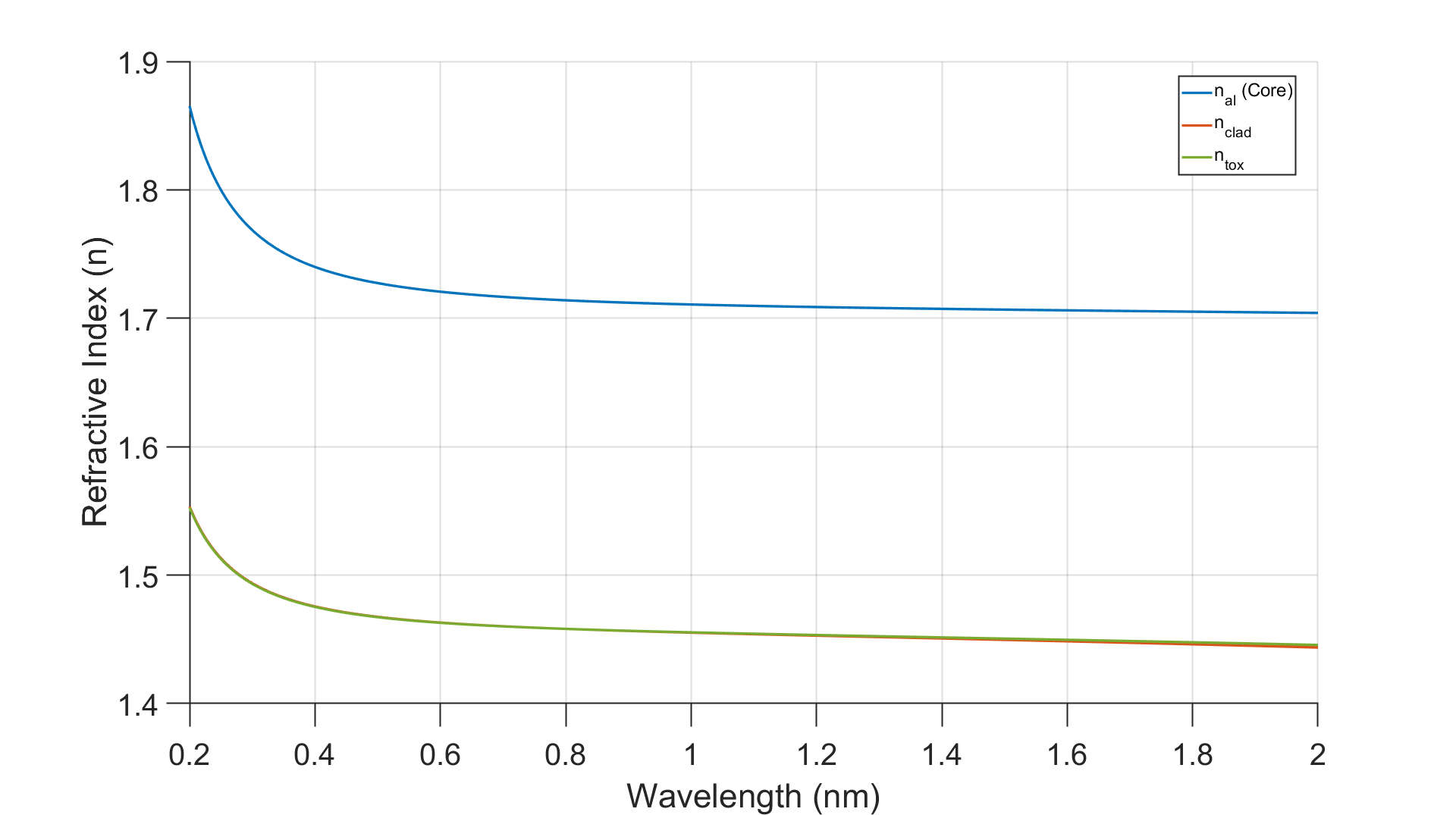}
        \caption{Measured refractive indices of the core (Al$_2$O$_3$) and surrounding materials.}
        \label{fig:refractiveindex}
    \end{subfigure}
    \caption{Platform layer stack and refractive index profile.}
    \label{fig:layer_and_index}
\end{figure}

Figure \ref{fig:layerstack} presents the layer stack of the platform. Aluminum oxide is 400 nm thick and the thicknesses of cladding and thermal oxide are 8 $\mu$m. The $Al_2O_3$ platform offers a moderate refractive index (n $>$ 1.7), which is well suited for AR/VR display engines. This index provides sufficient optical contrast for tight waveguide routing while reducing phase noise and scattering compared to higher-index alternatives. This moderate index supports beam-steering designs with a wider achievable field of view (FoV) while maintaining excellent UV-to-visible transparency, supporting high-efficiency red, green, and blue light propagation.

For the 400 nm thick waveguides, the fabricated sidewall angle of approximately 85$^\circ$ must also be considered, as it deviates from the ideal vertical profile. This non-vertical sidewall modifies the effective waveguide width and refractive index, which can influence mode confinement and increase scattering losses.

The gratings are designed to radiate light at specific angles; in this case, at $\sim$26$^\circ$. The chosen radiation angle corresponds to the smallest achievable grating period supported by the foundry design rules—200 nm for the grating at 452 nm. Attempts to design gratings with smaller periods were unsuccessful, as the grating teeth merged or collapsed during fabrication. The selected angle ensures well-defined grating features while maintaining the desired optical performance.

\begin{figure}[htbp]
    \centering
    \begin{minipage}{0.4\textwidth}
        \centering
        \includegraphics[width=\textwidth]{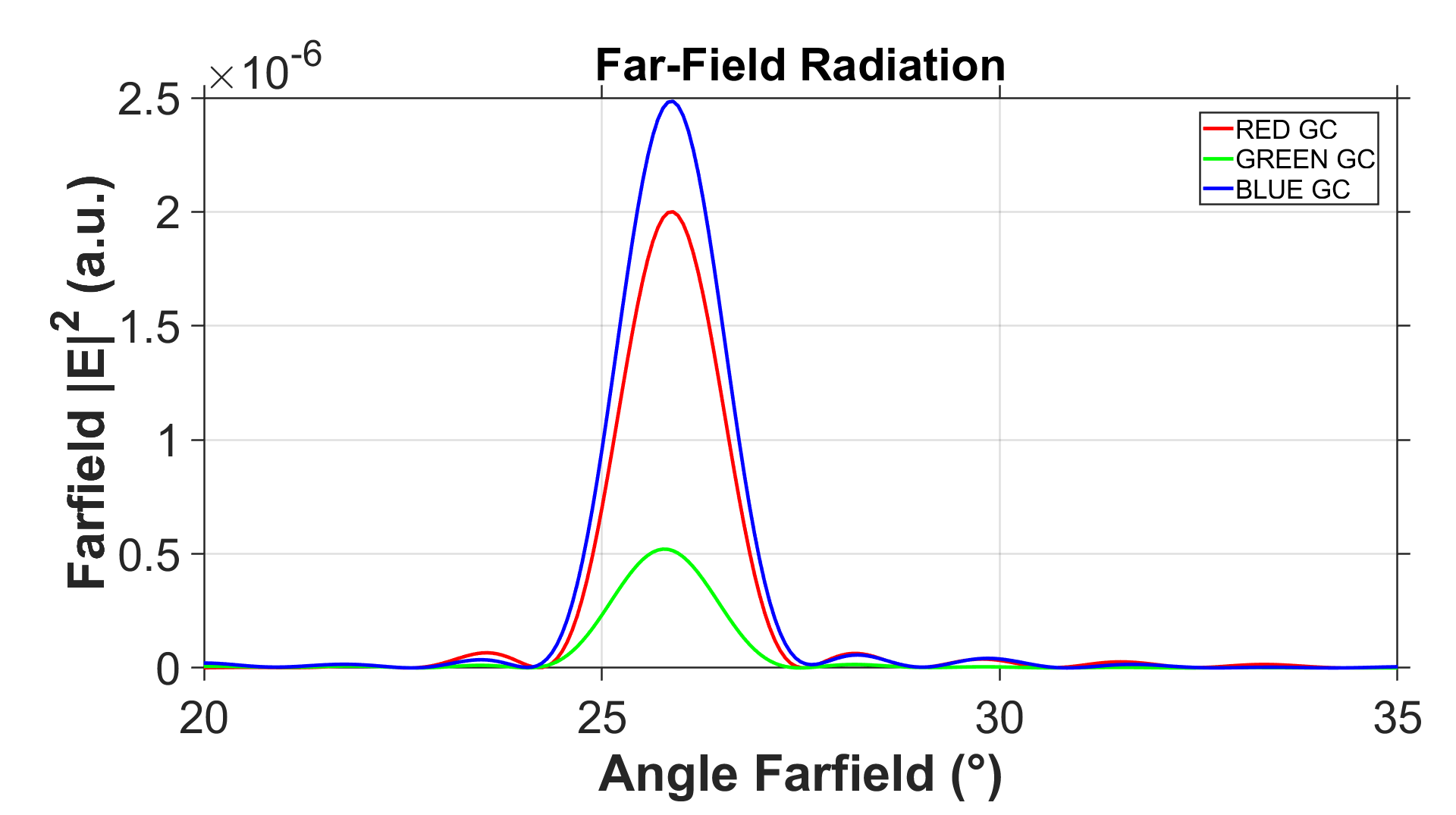}
        \text{(a)}
    \end{minipage}
    \hspace{0.05\textwidth}
    \begin{minipage}{0.4\textwidth}
        \centering
        \includegraphics[width=\textwidth]{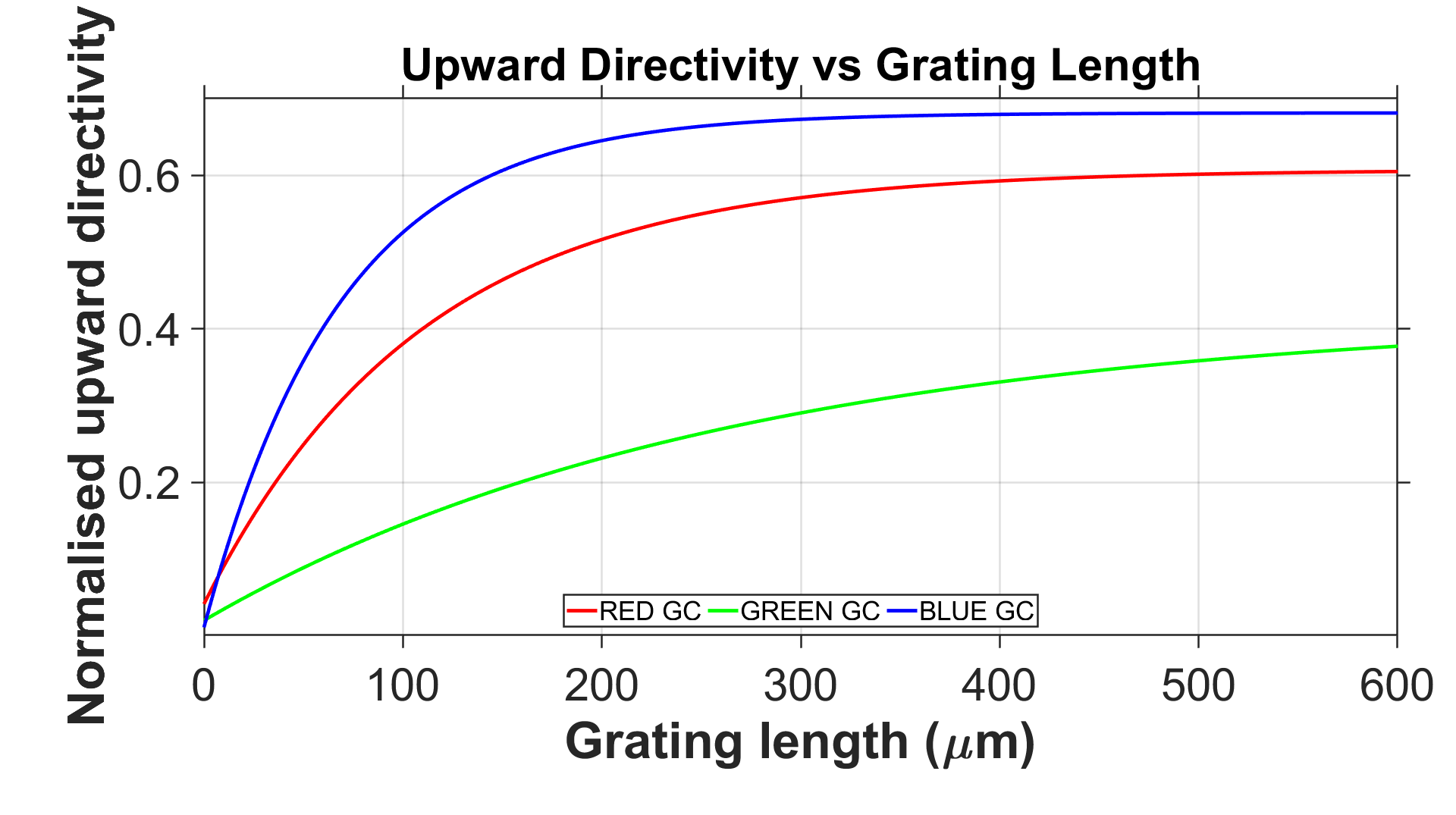}
        \text{(b)}
    \end{minipage}
    \caption{(a) Angle of emission from RGB gratings. (b) Upward directionality versus the length of RGB gratings.}
    \label{fig:length}
\end{figure}

The periods of the gratings are carefully adjusted to control the angle of light emission: 200 nm for blue light, 470 nm for green light, and 570 nm for red light, and the gratings are fully etched (see Fig.~\ref{fig:rgb_pixel_concept}(a)). In addition, the gratings are designed to achieve approximately 40--50\% upward directionality. This is accomplished by making the gratings sufficiently long, compensating for the low index contrast of the material (Fig.~\ref{fig:length}).

Mach--Zehnder modulators (MZMs), incorporating multimode interference (MMI) couplers for power splitting and recombination, are employed to control beam intensity and color. Phase tuning within the interferometer is achieved through thermo-optic modulation, which leverages the temperature dependence of the refractive index in Al$_2$O$_3$. By inducing a controlled phase shift between the two interferometer arms, the interference condition at the output is modified, enabling continuous intensity modulation and precise color control.

Thermo-optic tuning was selected due to its fabrication simplicity and compatibility with the current foundry process. Although this approach offers lower switching speeds compared to electro-optic modulation, it provides stable, reproducible, and low-complexity phase control well suited for proof-of-concept RGB pixel demonstrations.

\begin{figure}[htbp]
    \centering
    \includegraphics[width=0.45\textwidth]{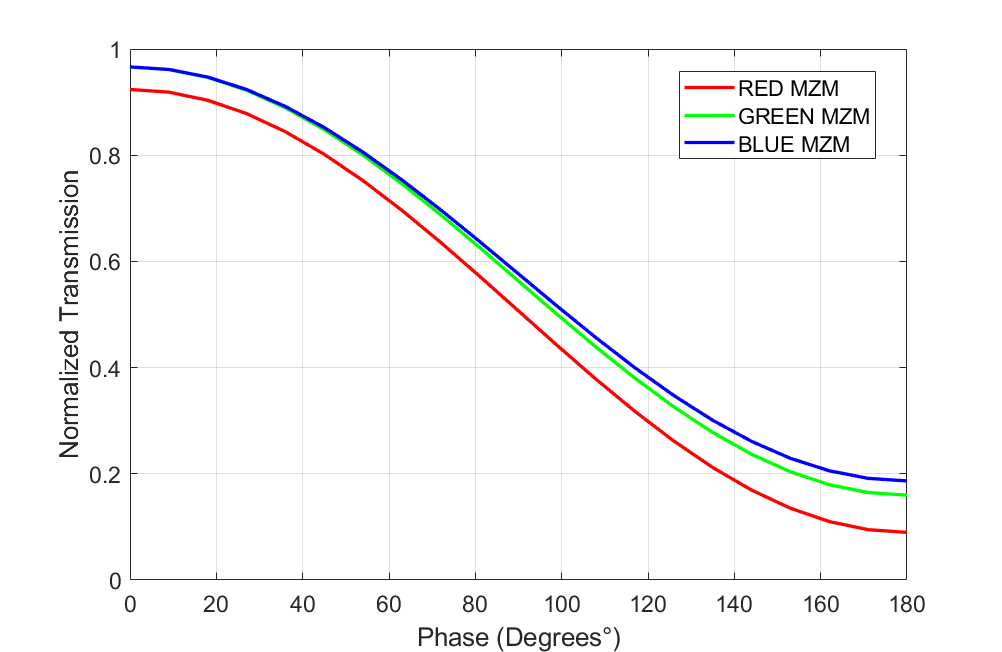}
    \caption{Phase versus amplitude change in the Mach--Zehnder modulator.}
    \label{fig:mzm_modulation}
\end{figure}

The graph in Figure \ref{fig:mzm_modulation} illustrates the relationship between phase and amplitude changes in the output port of the MMI. It shows how varying the phase difference affects the amplitude of the transmitted light, demonstrating the modulator’s ability to finely tune the light intensity and color output. The MMIs are designed specifically for each target wavelength to ensure optimal performance.

\begin{figure}[htbp]
    \centering
    \includegraphics[width=0.7\linewidth]{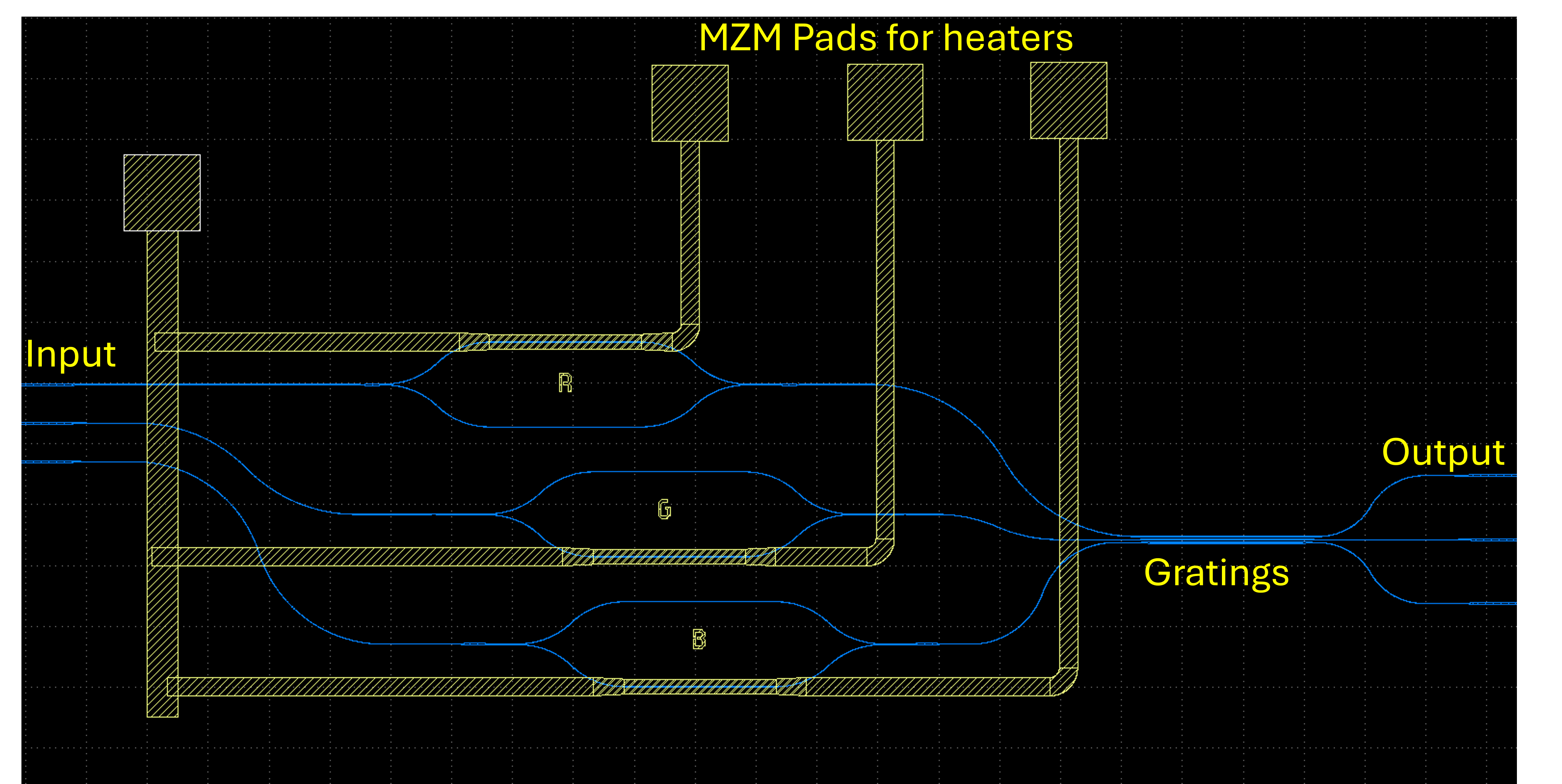}
    \caption{Full mask layout of the fabricated system.}
    \label{fig:mask_layout}
\end{figure}

The full mask layout for the system to be fabricated is shown in Figure~\ref{fig:mask_layout}. The input section is designed to match the standard fiber array pitch of 127~$\mu$m. The output at the chip edge also emits an RGB signal via edge-fire emission, which can be used for spectral analysis and later compared with the grating-based emission. The overall device footprint is 4 mm $\times$ 1 mm, making it a compact combiner suitable for augmented reality (AR) and virtual reality (VR) display systems. In future iterations, the lasers can be integrated on-chip to enable a fully integrated light engine.

For the next iteration, we will replace the Mach-Zehnder modulators (MZMs) with directional couplers (DCs) to reduce losses and reflections. Additionally, modifications to the grating layer stack will be made to improve upward directionality while maintaining compact grating dimensions. Ongoing research is focused on developing RGB on-chip combiners on different material platforms which can potentially enable a more compact design and facilitate higher-resolution integration of the gratings.

\section*{Fabrication}

Al$_2$O$_3$ layers with a thickness of approximately 400 nm were deposited using an AJA ATC 15000 RF reactive co-sputtering system on 100 mm diameter silicon wafers with an 8 $\mu$m thick thermal oxide layer.

The slab propagation losses were measured using a prism coupler setup (Metricon 2010M) equipped with a fiber loss measurement tool. Experimental measurements confirm the low-loss characteristics of the sputtered films at 377 nm wavelength for slab light propagation, with an attenuation of $0.40 \pm 0.05$~dB/cm measured for a 70~nm thick slab layer \cite{bonneville2024}.

The thickness and refractive index were measured using variable angle spectroscopic ellipsometry (VASE) with a Woollam M-2000UI ellipsometer on a similar Al$_2$O$_3$ layer deposited on a bare silicon substrate. At 452 nm, the propagation losses are $1.72 \pm 0.32$~dB/cm (TE) and $1.85 \pm 0.21$~dB/cm (TM).

\begin{figure}[h]
    \centering
    \includegraphics[width=0.42\textwidth]{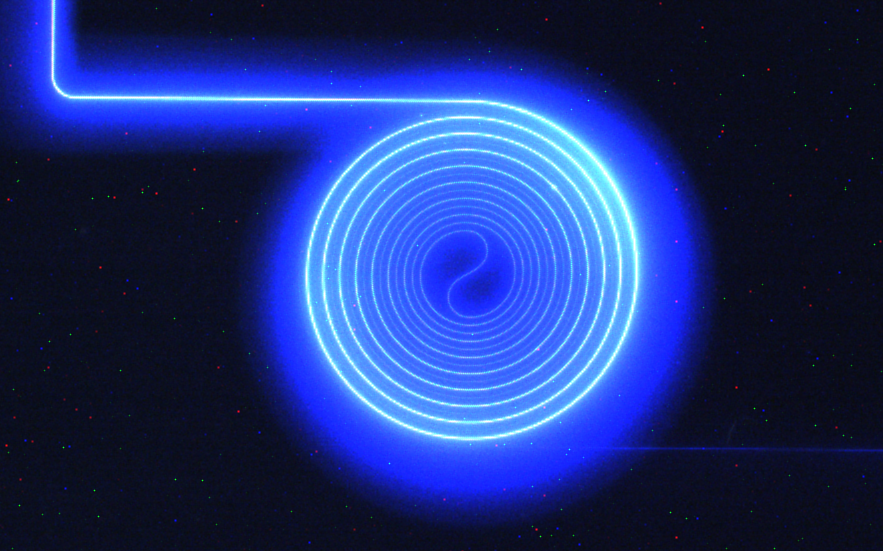}
    \caption{Blue/UV light at 369 nm propagating in fabricated spiral waveguides.}
    \label{fig:spiral}
\end{figure}

Aluminum oxide (Al$_2$O$_3$) waveguides exhibit low propagation loss in the blue spectral region and maintain good transparency down to the UV. This wide spectral range enables improved color gamut control, supporting efficient manipulation of red, green, and blue channels in integrated photonic devices.

\begin{figure}[htbp]
    \centering
    \includegraphics[width=0.45\textwidth]{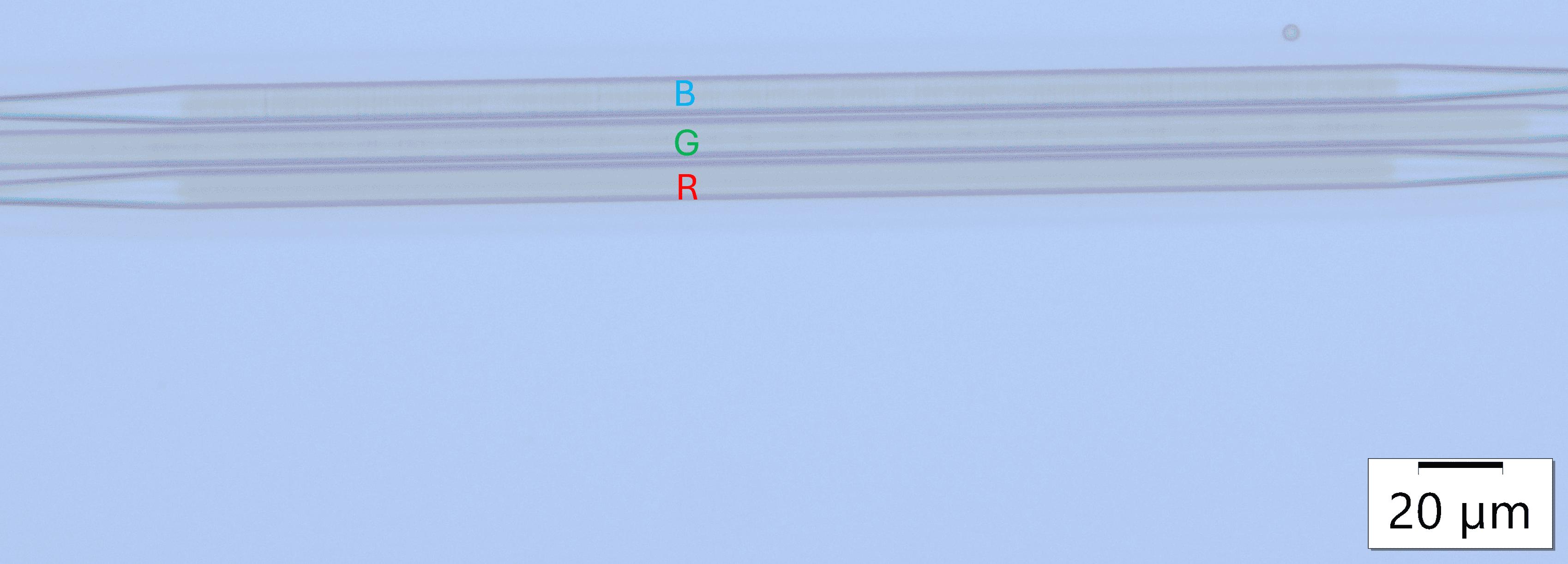}
    \caption{Microscope image of fabricated RGB grating structures.}
    \label{fig:fab_gratings}
\end{figure}

The microscope image of fabricated gratings is shown in Figure \ref{fig:fab_gratings}.

\section*{Experimental Results}

To demonstrate the operation of the gratings in the visible wavelength range on the aluminum oxide platform, a high-power Fianium supercontinuum laser was used as the input source. The broadband output enabled generation of a well-defined visible spectrum at the grating outputs. 

\begin{figure}[htbp]
    \centering
    \includegraphics[width=0.42\textwidth]{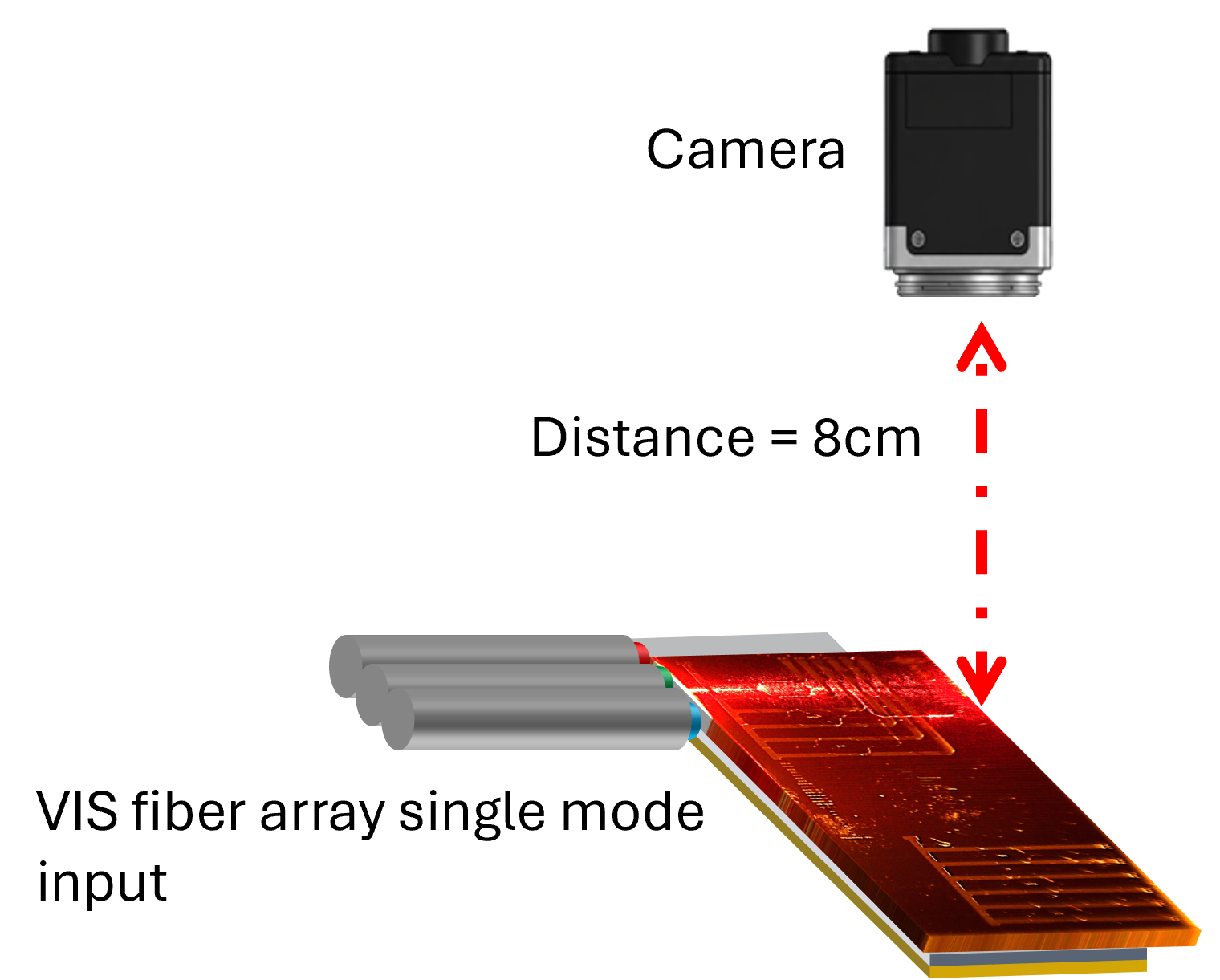}
    \caption{Simple setup to characterize grating emission.}
    \label{fig:experimental}
\end{figure}

A simple experimental setup was implemented to characterize the beams and their emission angles using a CCD camera. For the initial measurements, a multimode broadband optical fiber was used to couple the supercontinuum light into the chip. For subsequent RGB experiments, a three-channel single-mode fiber array for visible wavelengths, with a pitch of 127 $\mu$m, was employed to enable precise coupling to individual waveguides (Figure \ref{fig:experimental}).

\begin{figure}[htbp]
    \centering
    \includegraphics[width=0.35\textwidth]{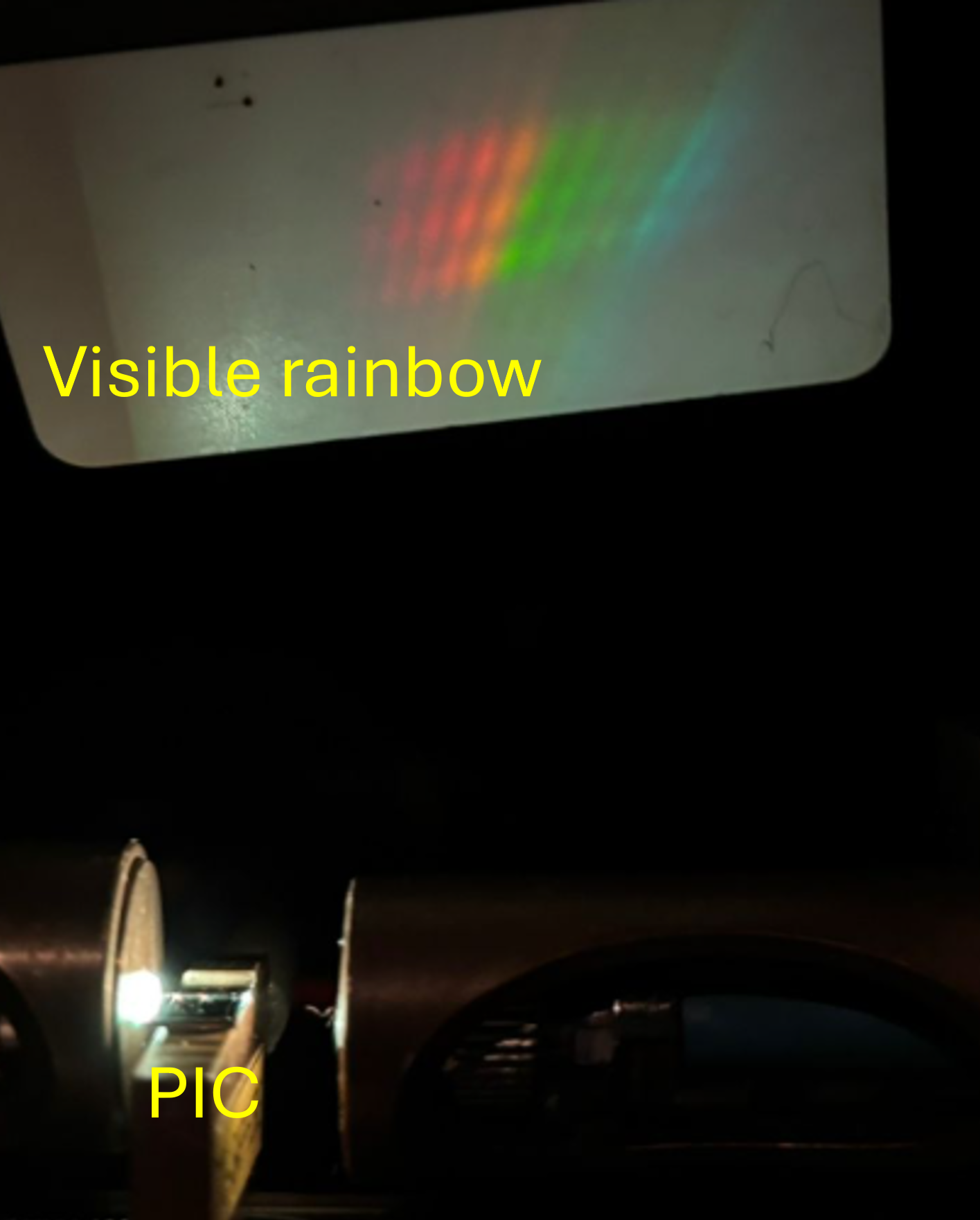}
    \caption{Visible rainbow pattern emitted from the test gratings, captured using the supercontinuum laser source.}
    \label{fig:grating_rainbow}
\end{figure}

Figure \ref{fig:grating_rainbow} shows a close-up of light coupling into the photonic chip using the Fianium laser, along with the resulting rainbow patterns emitted from the grating structures.

The CCD camera was positioned 8 mm from the grating emission region, and including the internal housing offset, the total distance from the photonic integrated circuit was 16 mm. The recorded intensity profile on the CCD sensor was used to determine the spatial separation between diffraction/interference peaks in pixel units. Using the calibrated pixel-to-length conversion, the physical separation between peaks was obtained.

Using the known emission wavelengths of the Fianium supercontinuum laser, the angular distribution of the diffracted beams was calculated using geometric triangulation. These angular measurements are critical for quantifying angular spread for visible wavelengths, which is an essential parameter in AR/VR optical system design, particularly for waveguide coupling and field-of-view optimization.

The fabricated test device exhibits an angular deviation with a root-mean-square error (RMSE) of $2.46^\circ$, corresponding to an approximate $\pm 3^\circ$ mismatch between experimental and simulated results. These differences are attributed to fabrication tolerances in the grating periods and etch depths, which can slightly alter the diffraction angles. For the designed test grating period of 0.52~$\mu$m, this corresponds to an effective fabrication variation of approximately $\pm 53$~nm on this aluminum oxide platform.

\begin{figure}[htbp]
    \centering
    \begin{subfigure}[b]{0.52\textwidth}
        \centering
        \includegraphics[width=\textwidth]{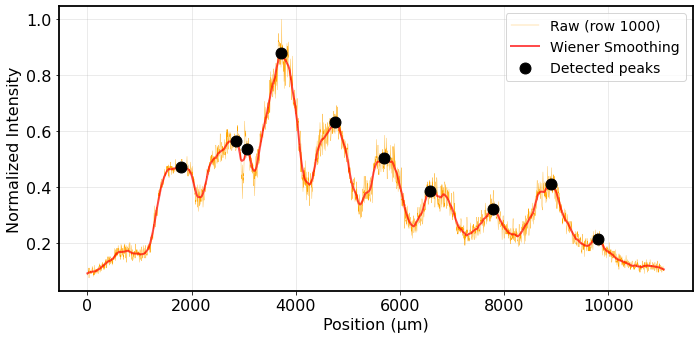}
        \caption{}
        \label{fig:rainbow_peaks}
    \end{subfigure}
    \hfill
    \begin{subfigure}[b]{0.45\textwidth}
        \centering
        \includegraphics[width=\textwidth]{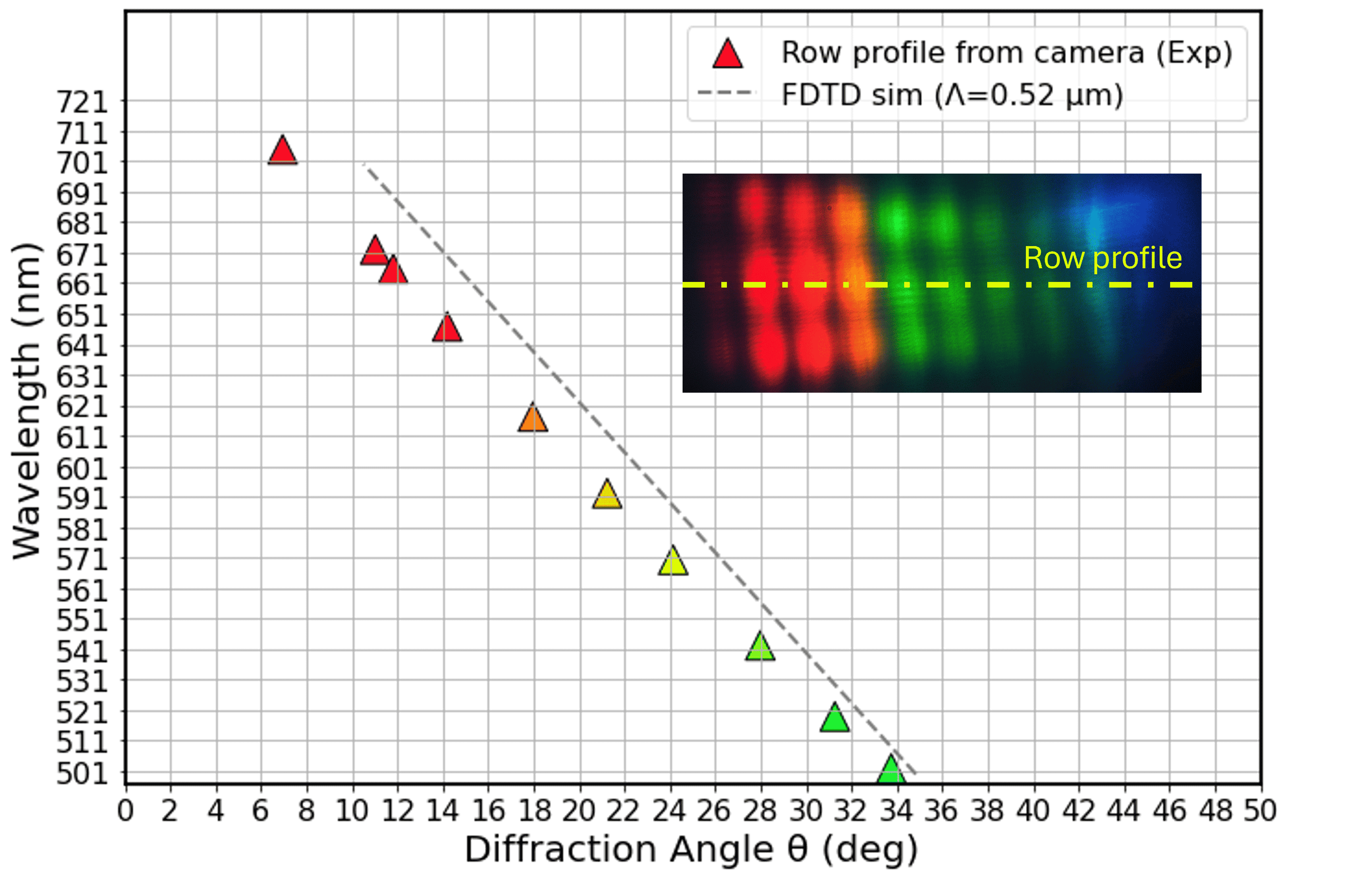}
        \caption{}
        \label{fig:angles_comparison}
    \end{subfigure}
    \caption{(a) Denoised (Wiener smoothing) rainbow diffraction peaks used for characterization. 
(b) Comparison of FDTD-simulated and experimentally measured emission angles.}
    \label{fig:angle_analysis}
\end{figure}

\subsection*{Additive color mixing and modulation}

\begin{figure}[htbp]
\centering
\rotatebox{90}{%
    \includegraphics[width=0.4\textwidth]{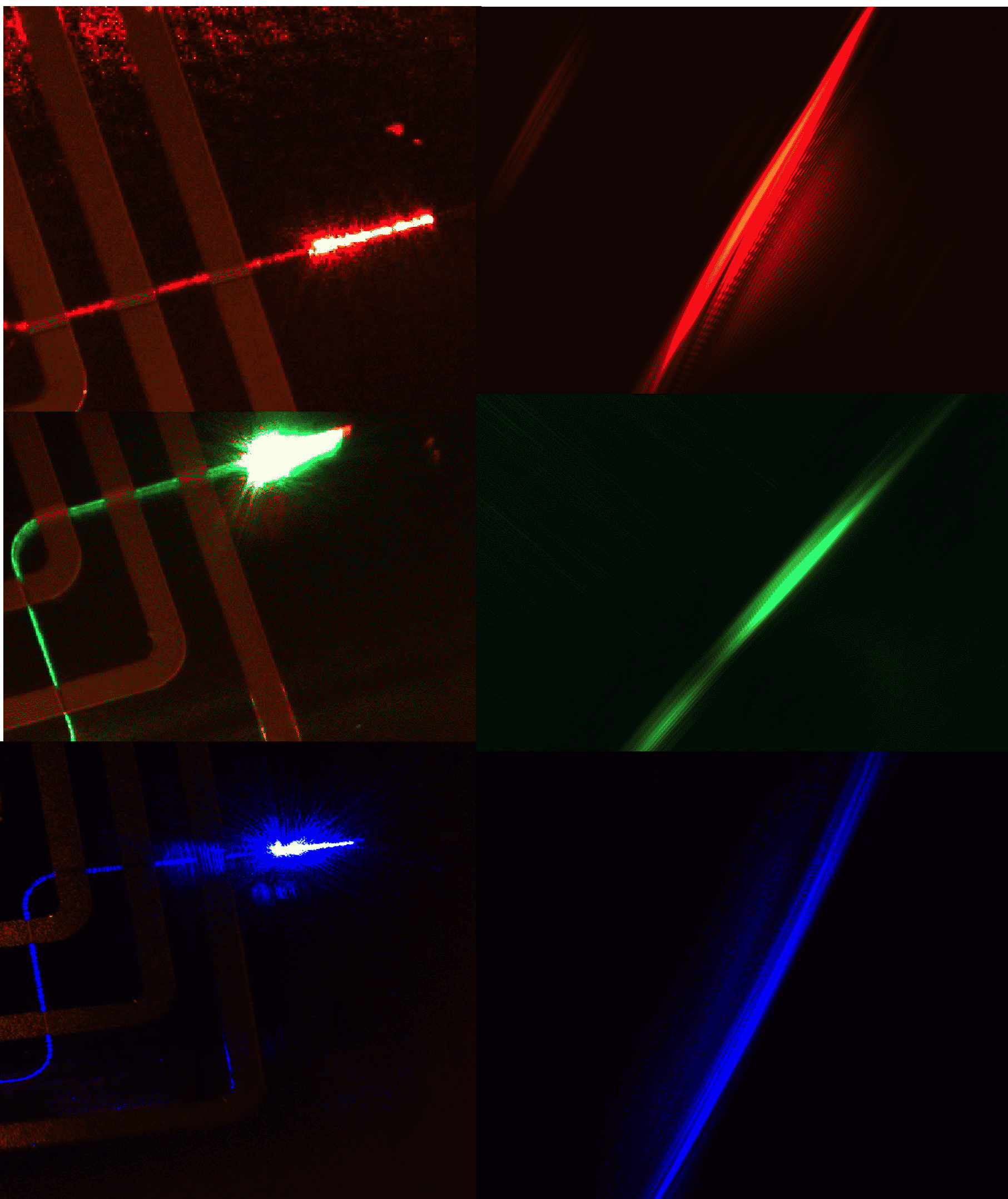}
}
\caption{Microscope image of the RGB combiner showing individual gratings and waveguides.}
\label{fig:RGB}
\end{figure}

Figure \ref{fig:RGB} shows the individual gratings emitting directly to a camera sensor. Each color—red, green, and blue—is launched into separate input waveguides and then emitted out of plane through the corresponding grating structures. The resulting images show distinct beam spots for each wavelength, verifying that RGB beams are individually addressable and spatially separated.

\begin{figure}[htbp]
\centering

\begin{subfigure}[t]{0.58\textwidth}
    \centering
    \includegraphics[width=\linewidth]{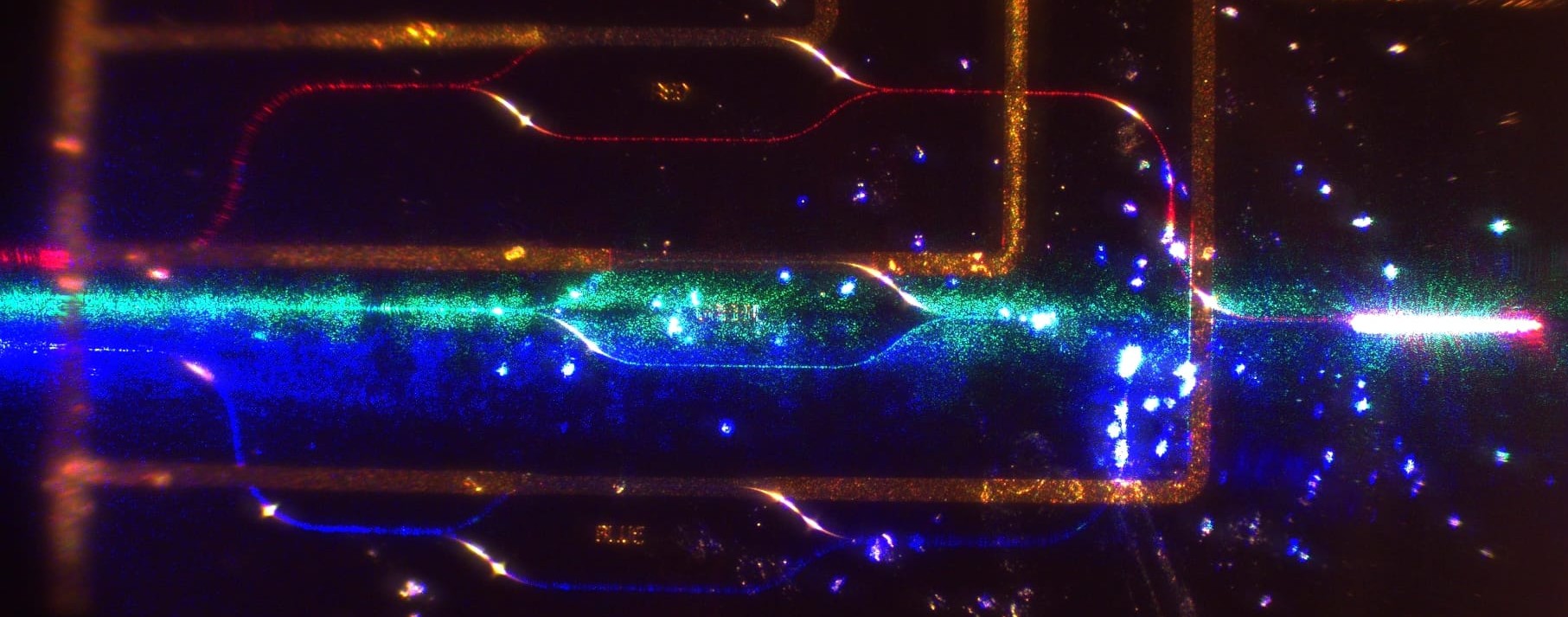}
\end{subfigure}\hfill
\begin{subfigure}[t]{0.38\textwidth}
    \centering
    \includegraphics[width=\linewidth]{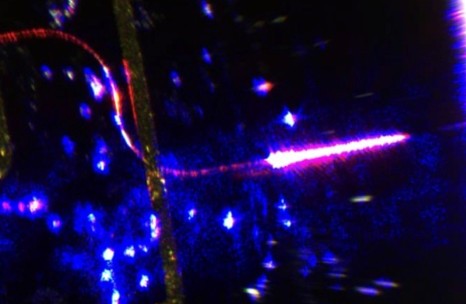}
\end{subfigure}

\caption{Microscope images showing simultaneous RGB coupling from the fiber array (left) and the resulting combined white light in the grating region. Turning off the green laser results in magenta output produced by red and blue additive mixing.}
\label{fig:rgb_white_light}
\end{figure}

Figure \ref{fig:rgb_white_light} shows the simultaneous coupling of light from the fiber array into the waveguides. At the output, white light is produced by the combined emission from the RGB gratings. By adjusting the input power in each channel, different colors can be generated; in this case, attenuating the green input produces a magenta output from red and blue additive mixing (Figure \ref{fig:rgb_white_light}b).

\begin{figure}[htbp]
\centering

\begin{subfigure}[t]{0.44\textwidth}
    \centering
    \includegraphics[width=\linewidth]{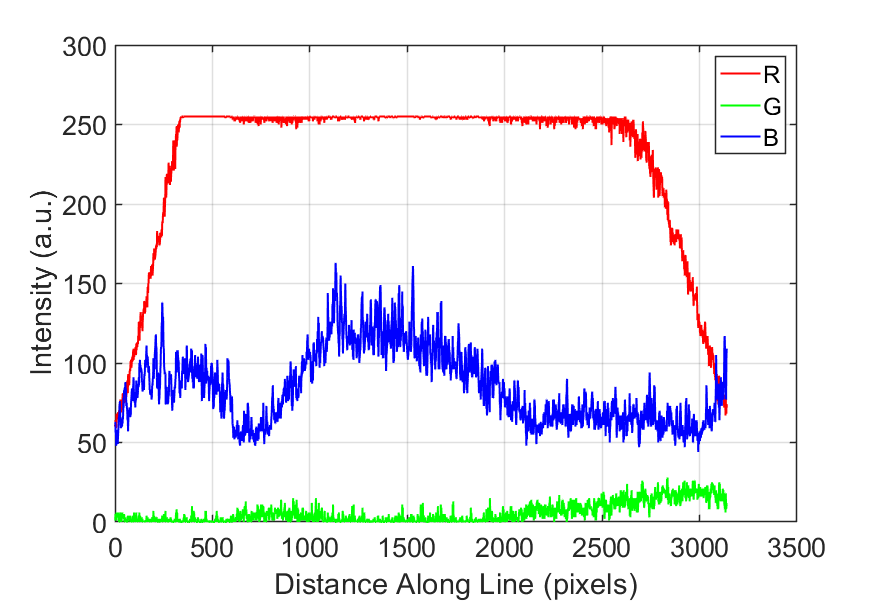}
    \caption{}
\end{subfigure}\hfill
\begin{subfigure}[t]{0.04\textwidth}
    \centering
    \includegraphics[width=\linewidth]{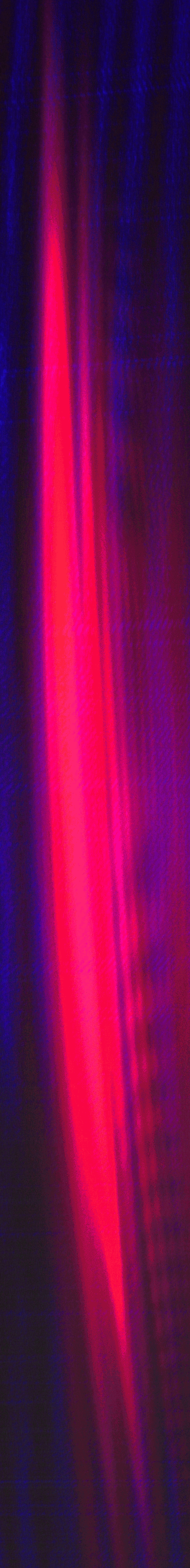}
    \caption{}
\end{subfigure}\hfill
\begin{subfigure}[t]{0.44\textwidth}
    \centering
    \includegraphics[width=\linewidth]{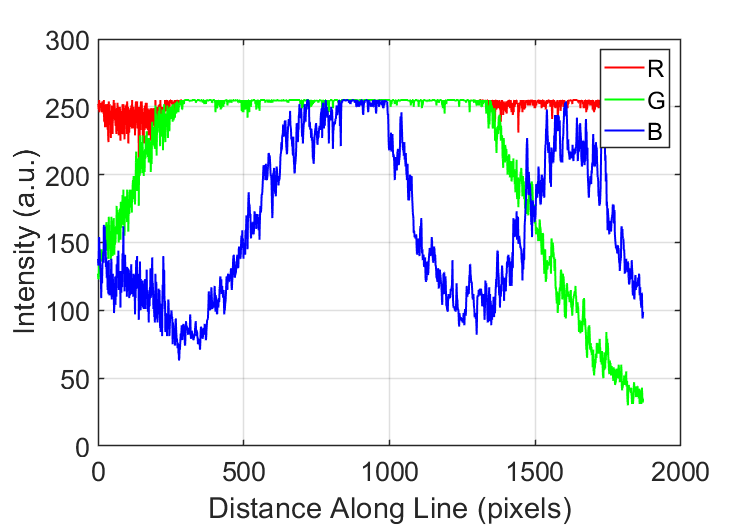}
    \caption{}
\end{subfigure}\hfill
\begin{subfigure}[t]{0.07\textwidth}
    \centering
    \includegraphics[width=\linewidth]{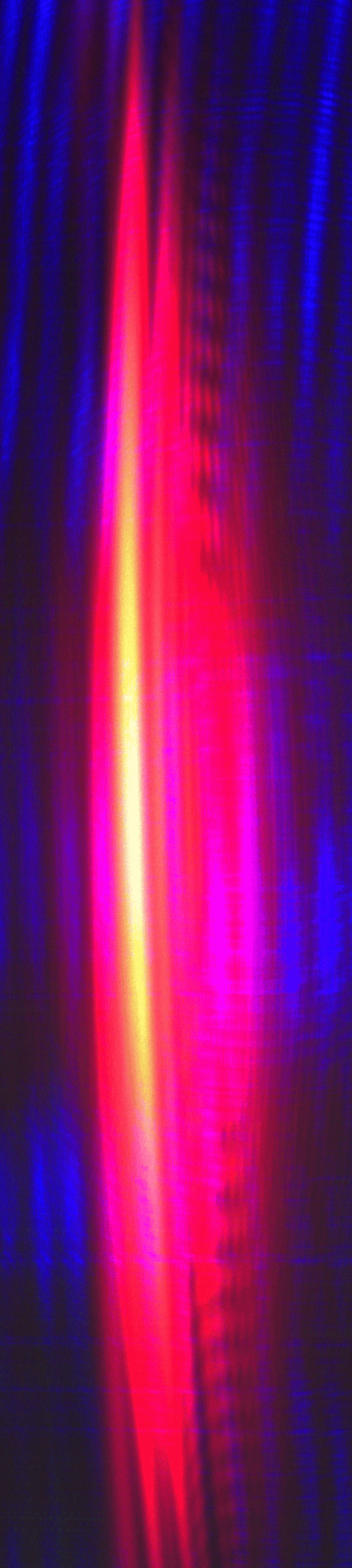}
    \caption{}
\end{subfigure}

\caption{Demonstration of additive RGB color mixing in the far field. 
(a) Line intensity profile extracted from the far field image when only the red and blue channels are active (green laser off). The profile shows spatially separated red and blue contributions without formation of a white central region. 
(b) Corresponding far field emission image showing the magenta-dominated beam profile resulting from red and blue overlap. 
(c) Line intensity profile when the green channel is enabled, showing the spatial overlap of red, green, and blue intensity distributions. 
(d) Far field emission image illustrating additive color mixing, where the overlap of the three channels produces a white central region.}

\label{fig:mixing}
\end{figure}

Figure \ref{fig:mixing} shows the far field emission from the three grating emitters. In the absence of green excitation (top row), only red and blue contributions are observed, resulting in a magenta-dominated angular profile without formation of a white central spot. When green excitation is introduced (bottom row), spatial overlap of red, green, and blue angular lobes produces additive color mixing in the far field, yielding a white central region. The corresponding RGB intensity profiles confirm increased green contribution within the overlap region.

The central region saturates the CCD sensor, and thus the apparent white coloration cannot be interpreted quantitatively. The recorded RGB balance may be affected by sensor spectral response, channel cross-talk, and nonlinear processing within the imaging system. A calibrated spectrometer measurement of the combined emission was not available at this stage. Therefore, the white-light generation is interpreted qualitatively from the controlled RGB excitation and the verified spatial overlap of their far field emission.

\begin{figure}[h!]
    \centering
    
    \begin{subfigure}[t]{0.45\textwidth}
        \centering
        \includegraphics[width=\linewidth]{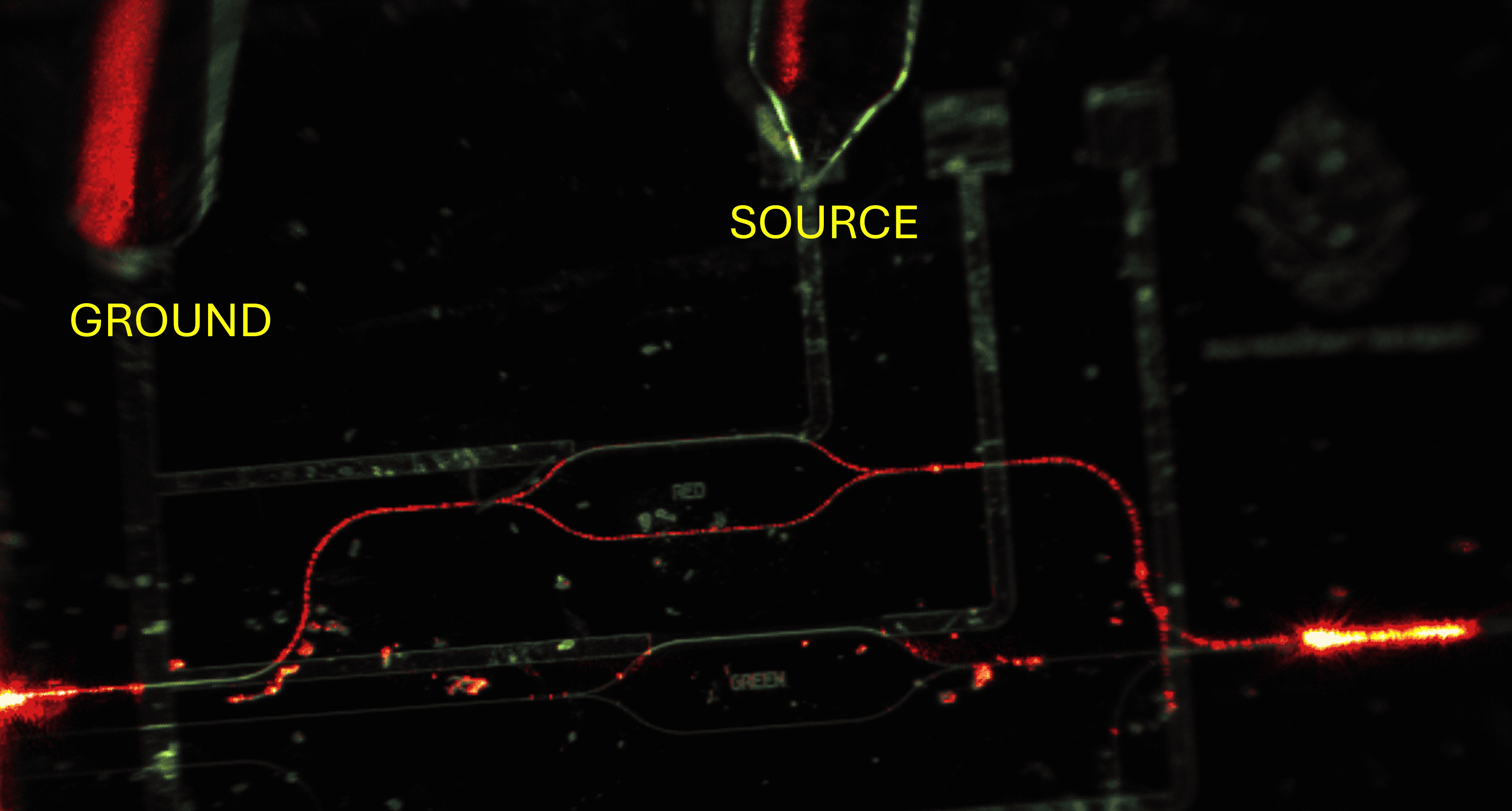}
        \caption{}
        \label{fig:red_no_bias}
    \end{subfigure}
    \hfill
    \begin{subfigure}[t]{0.45\textwidth}
        \centering
        \includegraphics[width=\linewidth]{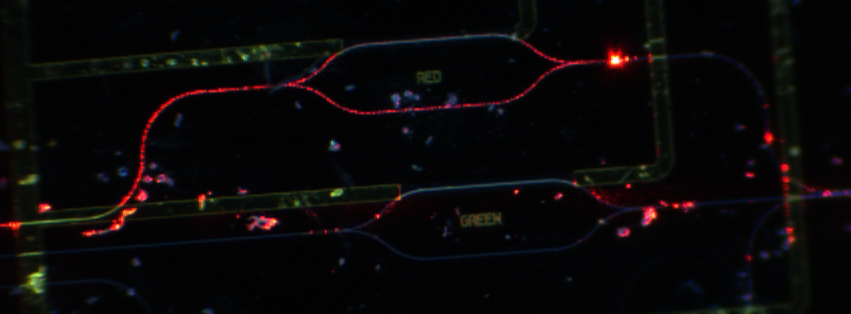}
        \caption{}
        \label{fig:modulation_red}
    \end{subfigure}
    
    \vspace{0.5cm}
    
    \begin{subfigure}[t]{0.6\textwidth}
        \centering
        \includegraphics[width=\linewidth]{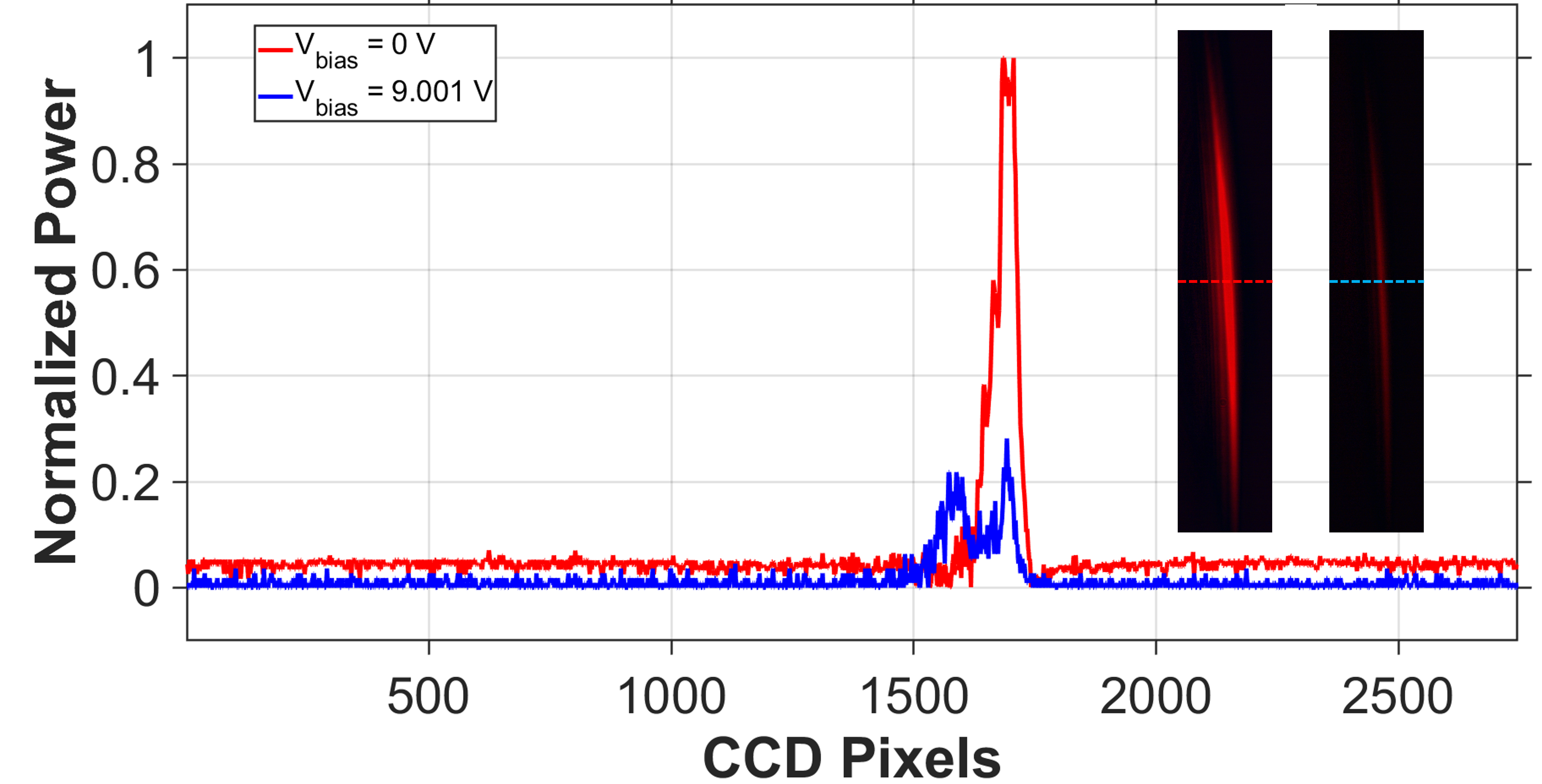}
        \caption{}
        \label{fig:red_suppression}
    \end{subfigure}
    
    \caption{Experimental demonstration of thermo-optic modulation in the red channel. 
    (a) Propagation without applied voltage. 
    (b) Suppression under applied bias. 
    (c) Measured extinction ratio showing suppression of the modulated signal.}
    \label{fig:red_modulation}
\end{figure}

Modulation of red light has been successfully demonstrated, with a peak suppression of 6.3~dB observed (Figure \ref{fig:red_modulation}). The red light coupled through the grating was projected onto a CCD camera while a bias voltage of 9.001 V was applied. Under this bias, the modulated signal clearly shows a suppressed peak, indicating effective modulation.

However, a residual stray-light component remains coupled into the grating, resulting in a secondary peak that is not part of the modulated signal. This stray component remains unaffected by the applied voltage.

Current work focuses on characterizing modulation at different wavelengths. Preliminary measurements indicate consistent modulation behavior in the blue channel. In the figure below, the modulation of the blue waveguides can be observed, along with the suppression of blue light in the far field.

\enlargethispage{2\baselineskip}

\begin{figure}[h!]
    \centering
    \includegraphics[width=0.41\textwidth]{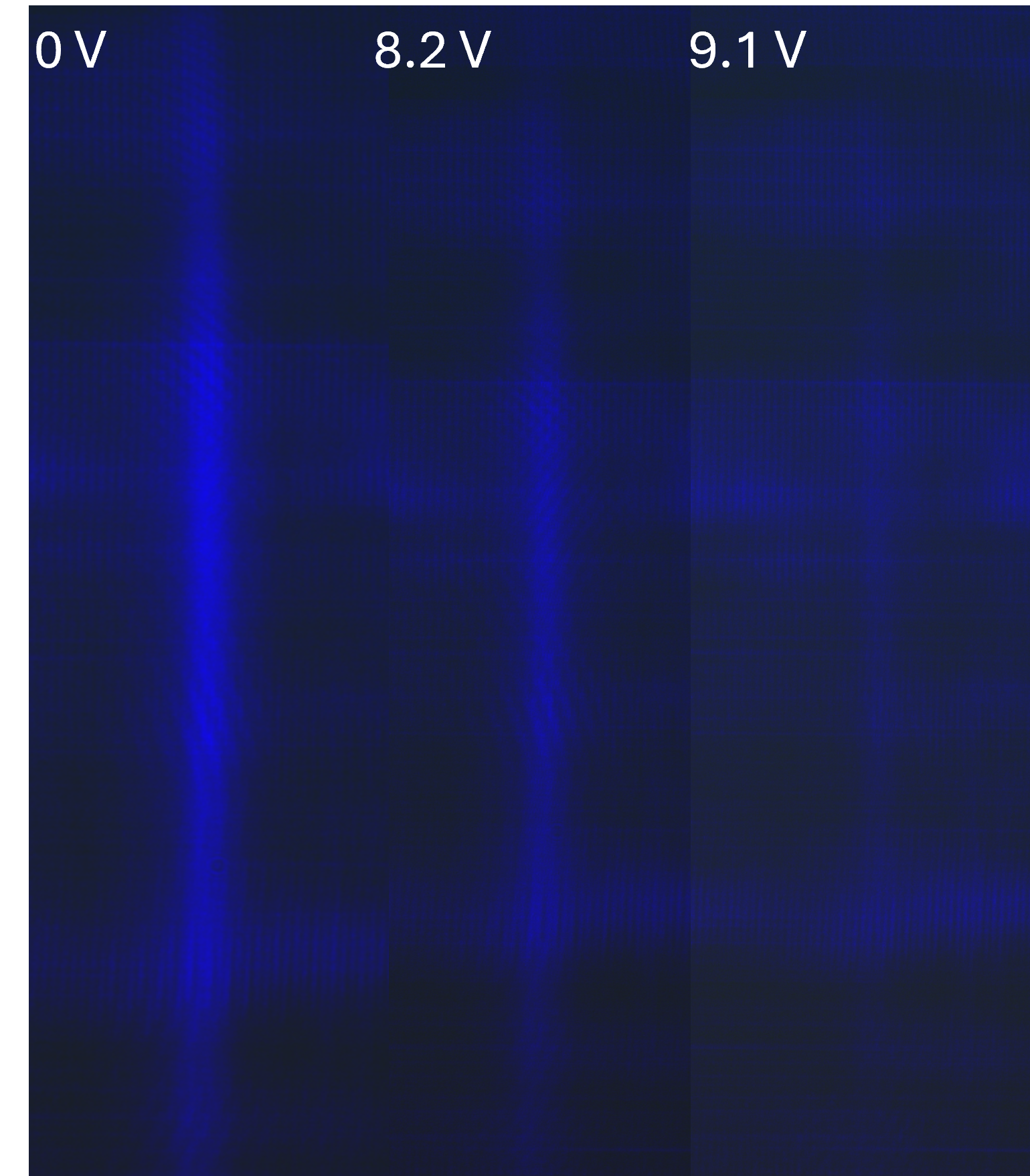}
    \caption{Blue (452 nm) far field extinction via applied voltage.}
    \label{fig:blue_extinction}
\end{figure}

After multiple characterization cycles, the chip surface became scratched, causing the blue light to appear murky. This effect is more pronounced at shorter wavelengths, which are more sensitive to surface roughness, as shown in Fig.~\ref{fig:blue_extinction}.

\clearpage

\section*{Outlook}

This paper demonstrates progress on the RGB beam combiner and outlines the current design approach. Future work will require a dedicated laboratory setup to properly project the beams in the desired direction and efficiently collimate the emitted light onto a CCD camera. Optical components such as focusing and cylindrical lenses will be employed to verify and optimize the beam profile.

Experimental validation of grating efficiency and emission angles remains essential, ensuring single-mode operation across all waveguides. Improved fabrication precision will be necessary to achieve finer grating resolution, enabling near-vertical emission while minimizing angular deviations introduced by fabrication limitations, including sidewall roughness and proximity-effect corrections. Such deviations can result in beam misalignment and reduced system performance.

Comprehensive experimental validation of the modulation functionality across all wavelengths is also required. Initial results demonstrate effective intensity control, with an extinction ratio of up to 6.3~dB for the red channel and promising suppression for blue. However, residual stray light and incomplete extinction indicate clear opportunities for further optimization.

Future design iterations will investigate the incorporation of slanted gratings to expand the field of view, potentially enabling glasses-free (autostereoscopic) 3D displays \cite{fattal2013}. Alternative modulation schemes will also be explored to further enhance overall system performance.
A multilayer waveguide architecture, in which the red, green, and blue channels are routed in vertically separated layers, will also be investigated. Such an approach could reduce in-plane routing complexity, minimize wavelength crosstalk, and enable independent optimization of grating couplers for each color. Vertical integration may also improve beam alignment and overall device compactness.

\begin{figure}[htbp]
    \centering
    \includegraphics[width=0.48\textwidth]{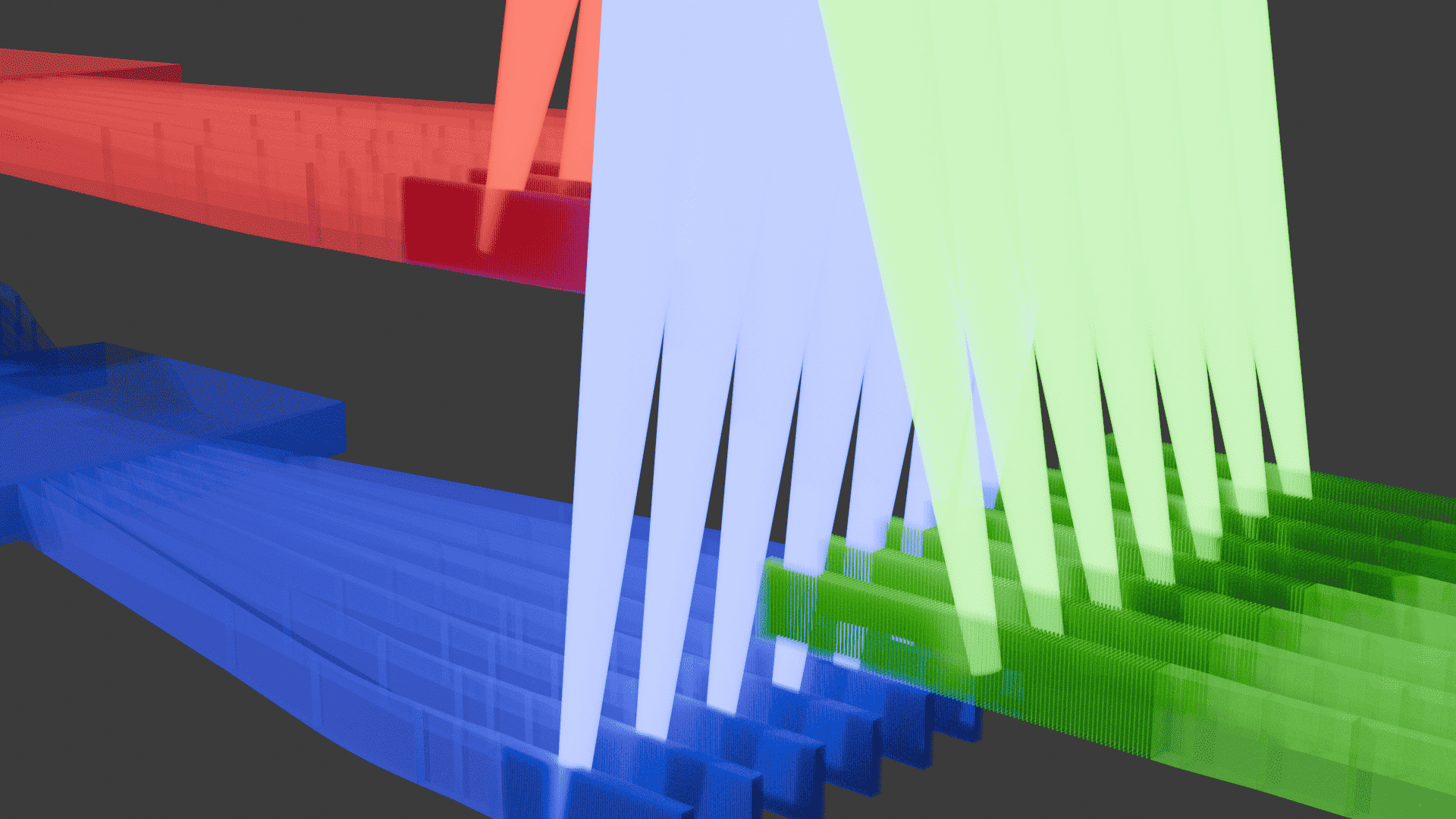}
    \caption{Multi-layer stack RGB combiner concept.}
    \label{fig:layers}
\end{figure}

A schematic illustration of the proposed multilayer configuration is shown in Fig.~\ref{fig:layers}, highlighting the vertically stacked waveguides and their respective grating emitters.

With these advancements, the RGB beam combiner shows strong potential for next-generation display technologies, offering compact, energy-efficient, and high-performance optical solutions.

\section*{Acknowledgments}

The author would like to acknowledge the use of laboratory resources provided by the IOS group, led by Prof.\ Dr.\ S.\ M.\ García Blanco at the University of Twente. Special thanks to Jeroen Korterik for assistance with laser integration. The author also gratefully acknowledges access to the MPW (multi-project wafer) service of Aluvia Photonics (\url{https://aluviaphotonics.com}) and the associated fabrication support, which enabled the prototyping and testing of the device. This work was supported in part by funding from PhotonDelta and the NextGEN High-Tech Semicon program.

\end{document}